\documentclass[ajl]{emulateapj}

\usepackage{graphicx,epsfig,natbib,color}
\usepackage{lscape}
\usepackage{graphicx}
\usepackage{epsfig}
\usepackage{natbib}
\usepackage[colorlinks, citecolor=blue]{hyperref}

\newcommand{\apjsub} {{ApJ submitted}}
\slugcomment{Accepted by ApJ}

\begin{document}
\title{Tracking the Evolution of A Coherent Magnetic Flux Rope Continuously from the Inner to the Outer Corona}

\author{X. Cheng$^{1,2}$, M. D. Ding$^{1,2}$, Y. Guo$^{1,2}$, J. Zhang$^{1,3}$, A. Vourlidas$^{4}$, Y. D. Liu$^{5}$, O. Olmedo$^{6}$, J. Q. Sun$^{1,2}$, \& C. Li$^{1,2}$}

\affil{$^1$ School of Astronomy and Space Science, Nanjing University, Nanjing 210093, China}\email{xincheng@nju.edu.cn}
\affil{$^2$ Key Laboratory for Modern Astronomy and Astrophysics (Nanjing University), Ministry of Education, Nanjing 210093, China}
\affil{$^3$ School of Physics, Astronomy and Computational Sciences, George Mason University, Fairfax, VA 22030, USA}
\affil{$^4$ Space Science Division, Naval Research Laboratory, Washington, DC 20375, USA}
\affil{$^5$ State Key Laboratory of Space Weather, National Space Science Center, Chinese Academy of Sciences, Beijing, China}
\affil{$^6$ Department of Physics, Institute for Astrophysics and Computational Sciences, The Catholic University of America, Washington, DC 20064, USA}

\begin{abstract}

The magnetic flux rope (MFR) is believed to be the underlying magnetic structure of coronal mass ejections (CMEs). However, it remains unclear how an MFR evolves into and forms the multi-component structure of a CME. \textbf{In this paper, we perform a comprehensive study of an extreme-ultraviolet (EUV) MFR eruption on 2013 May 22 by tracking its morphological evolution, studying its kinematics, and quantifying its thermal property.} As EUV brightenings begin, the MFR starts to rise slowly and shows helical threads winding around an axis. Meanwhile, cool filamentary materials descend spirally down to the chromosphere. These features provide direct observational evidence of intrinsically helical structure of the MFR. Through detailed kinematical analysis, we find that the MFR evolution experiences two distinct phases: a slow rise phase and an impulsive acceleration phase. We attribute the first phase to the magnetic reconnection within the quasi-separatrix-layers surrounding the MFR, and the much more energetic second phase to the fast magnetic reconnection underneath the MFR. We suggest that the transition between these two phases be caused by the torus instability. \textbf{Moreover, we identify that the MFR evolves smoothly into the outer corona and appears as a coherent structure within the white light CME volume.} The MFR in the outer corona was enveloped by bright fronts that originated from plasma pile-up in front of the expanding MFR. The fronts are also associated with the preceding sheath region followed the outmost MFR-driven shock.
\end{abstract}

\keywords{Sun: corona --- Sun: coronal mass ejections (CMEs) --- Sun: flares ---Sun: magnetic topology}
Online-only material: animations, color figures

\section{Introduction}
The magnetic flux rope (MFR) is a volumetric plasma structure with the magnetic field lines wrapping around a central axis. It has been invoked in various astrophysical contexts like the magnetotail of the Earth \citep{hughes87,moldwin91}, the ionosphere of Venus \citep{russell79}, the Nebula \citep{morris06}, and the black hole system \citep{meier01,yuan09}. In the solar atmosphere, the MFR is believed to be a fundamental magnetic structure \citep{forbes91,chenj96,titov99,chen11_review}, which can erupt from the Sun as a coronal mass ejection (CME). The CME subsequently propagates into the interplanetary space and takes on the form of a magnetic cloud \citep[with typical features of the rotation of magnetic field, decreasing solar wind speed, depressed proton temperature, and low plasma beta;][]{burlaga88,lepping90,liuy08,liuy10,liuying11}, driving geomagnetic storms and thus impacting the space environment around the Earth \citep{gosling93,zhang07_icme}.

\begin{figure*}
\center {\includegraphics[width=15cm]{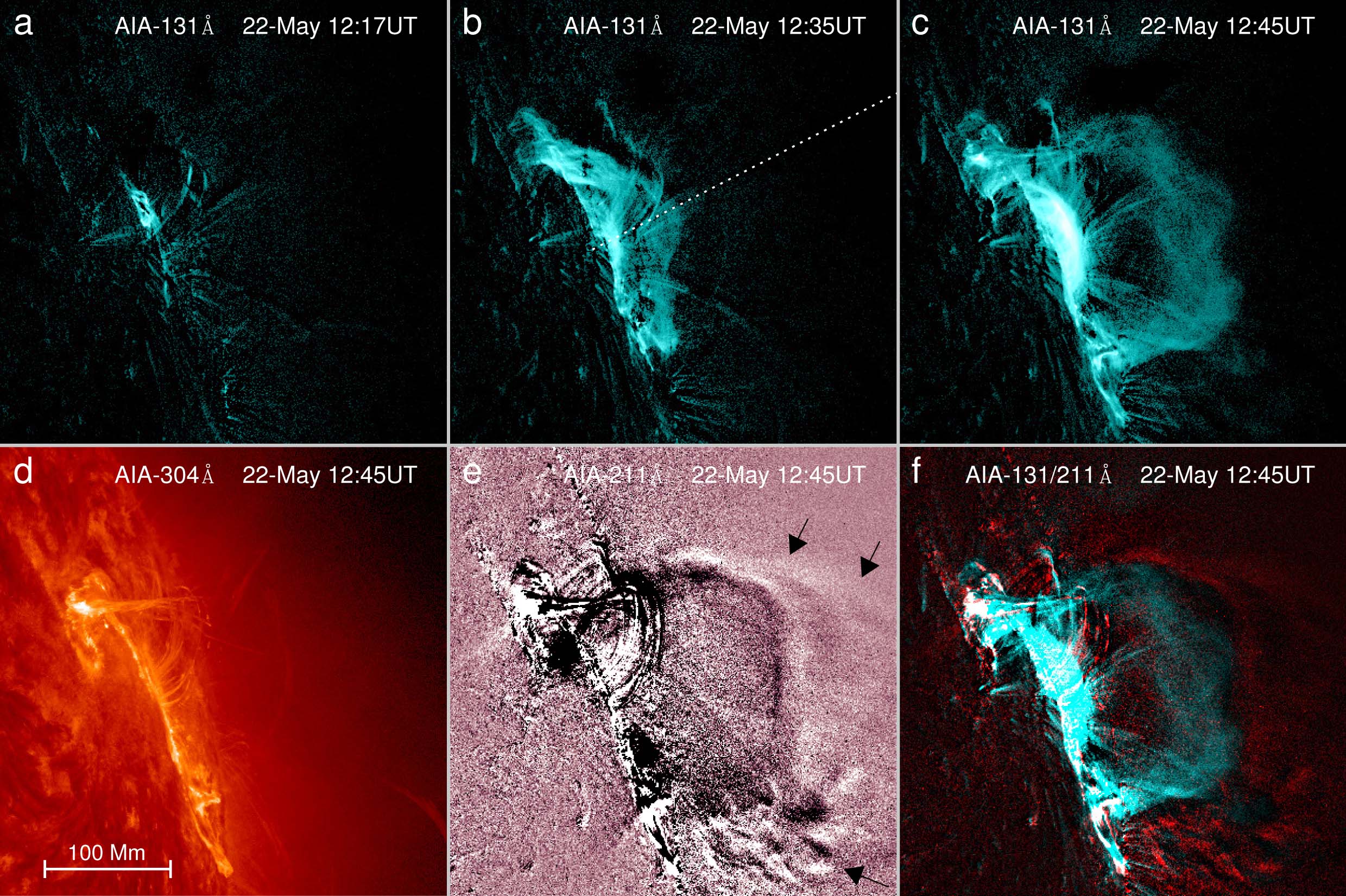}}
\caption{(a)--(c) The AIA 131 {\AA} ($\sim$0.4 MK and 11.0 MK) base difference images (the base image is taken at 12:00 UT) showing the evolution and fine structure of the MFR in the inner corona. The dashed line in panel (b) denotes the position of a slice at which two stack plots (Figure \ref{f2}(a, b)) are constructed. (d) The AIA 304 {\AA} ($\sim$0.05 MK) image displaying the cool filament material associated with the MFR. (e) The AIA 211 {\AA} ($\sim$2.0 MK) running difference image (the base image is taken at 12:43 UT) showing the CME front (black arrows). (f) Composite of the AIA 131 {\AA} (blue) and 211 {\AA} (red) difference images. }
(Animations this figure are available in the online journal.)
\label{f1}
\end{figure*}

Theoretically, MFRs can be formed in two ways: bodily flux emergence from below the photosphere or magnetic reconnection of sheared arcades in the corona. In the emergence model, a twisted MFR is assumed to exist below the photosphere and emerge into a pre-existing coronal potential field \citep{fan01,fan10,manchester04,magara06}. In the reconnection model, the imposed boundary movements such as converging and shearing motions of different polarities, rotation of sunspots, and magnetic flux cancellation, twist and stretch the initial potential field gradually, leading to magnetic reconnection \citep[e.g.,][]{amari11,aulanier10}. The magnetic reconnection between the sheared fields, known as the tether-cutting or breakout reconnection, is able to form the MFR during the eruption \citep{moore01,antiochos99,lynch08,karpen12}. Moreover, using observed photospheric vector magnetic field as the bottom boundary, the topology of MFR in the corona can be reconstructed through extrapolation of the nonlinear force-free field models \citep[e.g.,][]{yan01,canou09,guo10_filament,cheng10_reflare,cheng13_double,suyingna11,jiang13,inoue13}.

Assuming that the MFR is a line current,  \citet{forbes91} found that the MFR can suddenly lose equilibrium and erupt upwards explosively when ascending to a critical height. Other ideal magnetohydrodynamic (MHD) instabilities such as kink and torus instability are also able to initiate the eruption of the MFR. The ideal kink mode will develop nonlinearly if the twist of the MFR exceeds a threshold value \citep{torok04}. Torus instability refers to the expansion instability of a torus current. \citet{kliem06} showed that the torus instability would be triggered when the decline of the background field in the direction of the expansion of the MFR is sufficiently rapid. \citet{olmedo10} revealed that the threshold value for triggering torus instability depends on the geometrical circularity of the FR: the ratio between the MFR arc length above the photosphere and its circumference. \citet{demoulin10} further showed that the loss-of-equilibrium and torus instability actually refer to the same physical process and can be unified in the framework of ideal MHD.  

Owing to the importance of the role of MFRs in solar eruptions, researchers have been looking for observational evidence of MFRs. Through inspecting a sequence of vector magnetograms, \citet{okamoto08} found that two opposite polarity regions with vertically weak but horizontally strong magnetic field grew laterally and then narrowed in size. The orientations of the horizontal magnetic fields along the neutral line gradually changed from a normal polarity configuration to an inverse polarity one. The results imply that an MFR is likely emerging from below the photosphere. Based on a statistical study, \citet{canfield99} discovered that the morphology of active regions (ARs) usually appears as a forward or reversed sigmoid. The straight sections of the double J-shaped patterns in the middle of the sigmoid are often interpreted as evidence of MFR formation prior to the eruption \citep[e.g.,][]{mcKenzie08,green09,tripathi09,liur10,savcheva12a}. 

\begin{figure*}
\center {\includegraphics[width=15cm]{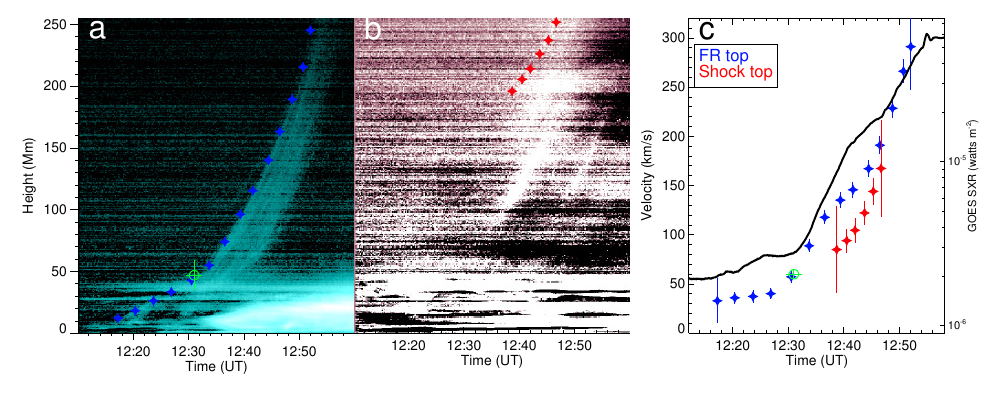}\vspace{-0.04\textwidth}}
\center {\includegraphics[width=14.23cm]{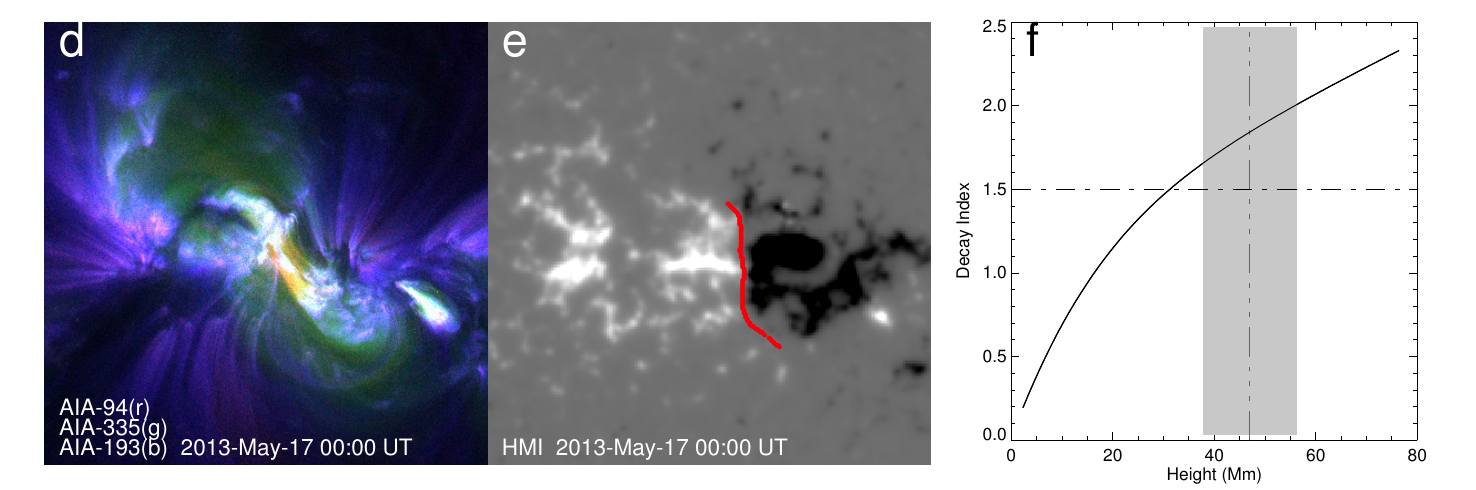}\hspace{-0.045\textwidth}}
\caption{(a) Stack plot of the 131 {\AA} ($\sim$0.4 MK and 11.0 MK) intensity along the slice in Figure \ref{f1}(b). Blue stars indicate the height measurements of the MFR above the solar surface, green circle denotes the onset time of the impulsive acceleration of the MFR with the horizontal (vertical) bar showing the uncertainty of the onset time (height). (b) Same as (a) but for the CME front. (c) Temporal evolutions of the MFR and CME front velocity in the inner corona. The back solid line shows the $GOES$ 1--8 {\AA} soft X-ray flux of the associated flare. Green circle has the same meaning as that in panel (a) but the vertical bar shows the uncertainty of the velocity at onset time. (d) Composite of the AIA 94 {\AA} ($\sim$7.0 MK), 335 {\AA} ($\sim$2.5 MK), and 193 {\AA} ($\sim$1.0 MK) images. (e) Line-of-sight magnetogram of the AR where the MFR originated. The red curve denotes the main neutral line. (f) Distribution of the decay index of the background field with height  over the main neutral line. The vertical dash-dotted line denotes the onset height of the impulsive acceleration of the MFR with the gray region showing the uncertainty. The horizontal dash line indicates the threshold value 1.5 for the torus instability.}

(An animation this figure is available in the online journal.)
\label{f2}
\end{figure*}

\textbf{Filaments are another proxy for the existence of MFRs in the corona as they are thought to arise from the collection of cool plasma in the dips of the MFR helical lines \citep{gibson04,mackay10,guo10_filament,suyingna11,suyingna12}. Filament channels are believed to be the body of the MFR \citep{low95_apj,guo98}, which often manifest as dark cavities in visible or EUV light when rotating to the solar limb \citep{gibson06_apj,gibson06_jgr,regnier11,berger12}. The ubiquitous spinning motions \citep{wangym10,lixing12},  as well as the appearance of the bright ring and ``lagomorphic" structure of linear polarization \citep{dove11,bak-steslicka13}, indicate that cavities contain helical structures.} Recently, \citet{zhang12} and \citet{cheng13_driver} reported the existence of EUV hot channels that appeared in the high temperature passbands of the Atmospheric Imaging Assembly (AIA) telescope \citep{lemen12} tens of minutes before the eruption. Once the impulsive acceleration phase starts, the hot channel erupts upwards and develops into a semicircular shape \citep[also see;][]{liur10,patsourakos13,lileping13}. Detailed morphology and kinematic analyses suggest that the hot channel is most likely to be the MFR and plays a critical role in forming and accelerating the CME in the inner corona \citep{cheng13_driver,cheng13_double,patsourakos12}.

Although many observational investigations on the MFR have been made in the past, direct observations of the helical magnetic field patten inside MFRs remains rare. In this paper, we address this issue by analyzing a well observed MFR eruption on 2013 May 22. We find that the appearance of the helical threads inside the hot channel and the spirally descending movement of the filament materials along the two legs of the channel, provide strong evidence of the existence of MFRs. Another question is the relationship between the MFR seen in the inner corona and the CME structural components in the white-light observations seen in the outer corona. This event clearly shows that the MFR evolves coherently into the outer corona, almost completely filling the CME cavity, and forms a compression region ahead including the pile-up of plasma and the shock wave in front of the MFR as suggested. The evolution of the MFR in the inner corona is showed in Section 2, followed by the discussion of the evolution in the outer corona in Section 3. The summary and discussions are given in Section 4.

\begin{figure*}
\center {\includegraphics[width=15cm]{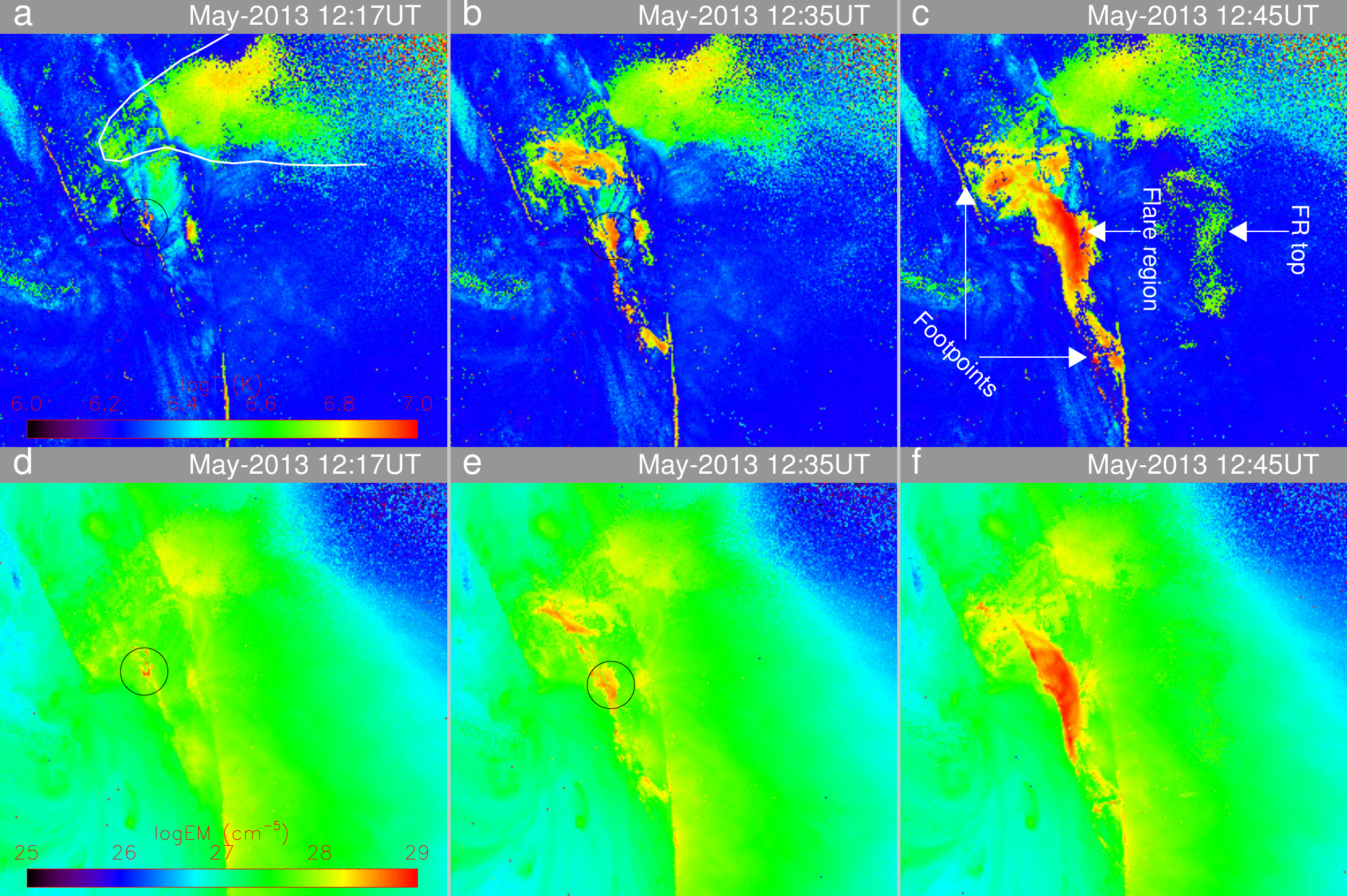}}\\
\caption{Average temperature (a--b) and total EM (d--f) maps of the 2013 May 22 CME at different instants. To calculate the average temperature and total EM, the DEM is integrated over the temperature range of 5.7 $\leq \log T \leq$ 7.3. The white contour in panel a is a region with $\overline{T}\geq$5.0 MK, which is caused by the previous eruption taking place at $\sim$08:00 UT.}

\label{f3}
\end{figure*}

\section{Evolution of MFR in the Inner Corona}

\subsection{Helical Features of MFR}
\textbf{At $\sim$12:10 UT, an EUV channel (elongated structure) reminiscent of a filament appeared in the AIA 131 \AA\ passband. No signatures were seen in the other AIA wavelengths. Therefore, the structure must have been very hot ($\sim$10MK). This is consistent with past detections of pre-eruption MFRs \citep{zhang12,cheng13_driver,patsourakos13}. The structure, which we will refer to as an MFR from now on, started to erupt from NOAA AR 11754 showing a sigmoidal structure that was located at the heliographic coordinates $\sim$N13W78.} The fine structures and detailed evolution of the MFR in the inner corona were well revealed by AIA on board \textit{Solar Dynamic Observatory} (\textit{SDO}), as shown in Figure \ref{f1}a--c, thanks to AIA's ability of high spatial resolution (1.2$\arcsec$), high temporal cadence (12 seconds), and multi-temperature coronal imaging (six EUV passbands). 

\textbf{During the eruption, while the two ends of the MFR were anchored on the chromosphere, most of the filament materials spirally descended into the chromosphere along the two legs of the MFR possibly due to gravity (Figure \ref{f1}d). The spiral movement is very similar to the spinning motion in coronal dark cavities as seen from the axis of the MFR \citep[e.g.,][]{wangym10,lixing12,liting13}.} At the same time, magnetic reconnection underneath the MFR, suggested by the EUV brightenings (Figure \ref{f1}a), heated the plasma and made the temperature in the MFR increase (Figure \ref{f1}b--c). It is well known that the plasma in the corona is frozen into the magnetic field due to high conductivity of the ionized corona \citep{priest00}, and the spatial distribution of the emission generally outlines the geometry of the magnetic field. Therefore, the helical threads inside the MFR, well imaged in the high temperature passbands of the AIA \citep[$\sim$10.0 MK at 131 {\AA} and $\sim$6.4 MK at 94 {\AA};][]{odwyer10,cheng11_fluxrope}, show the helicity of the MFR (Figure \ref{f1}c and attached movies).

\begin{figure*}
\vspace{-0.0\textwidth}
\center {\includegraphics[width=15cm]{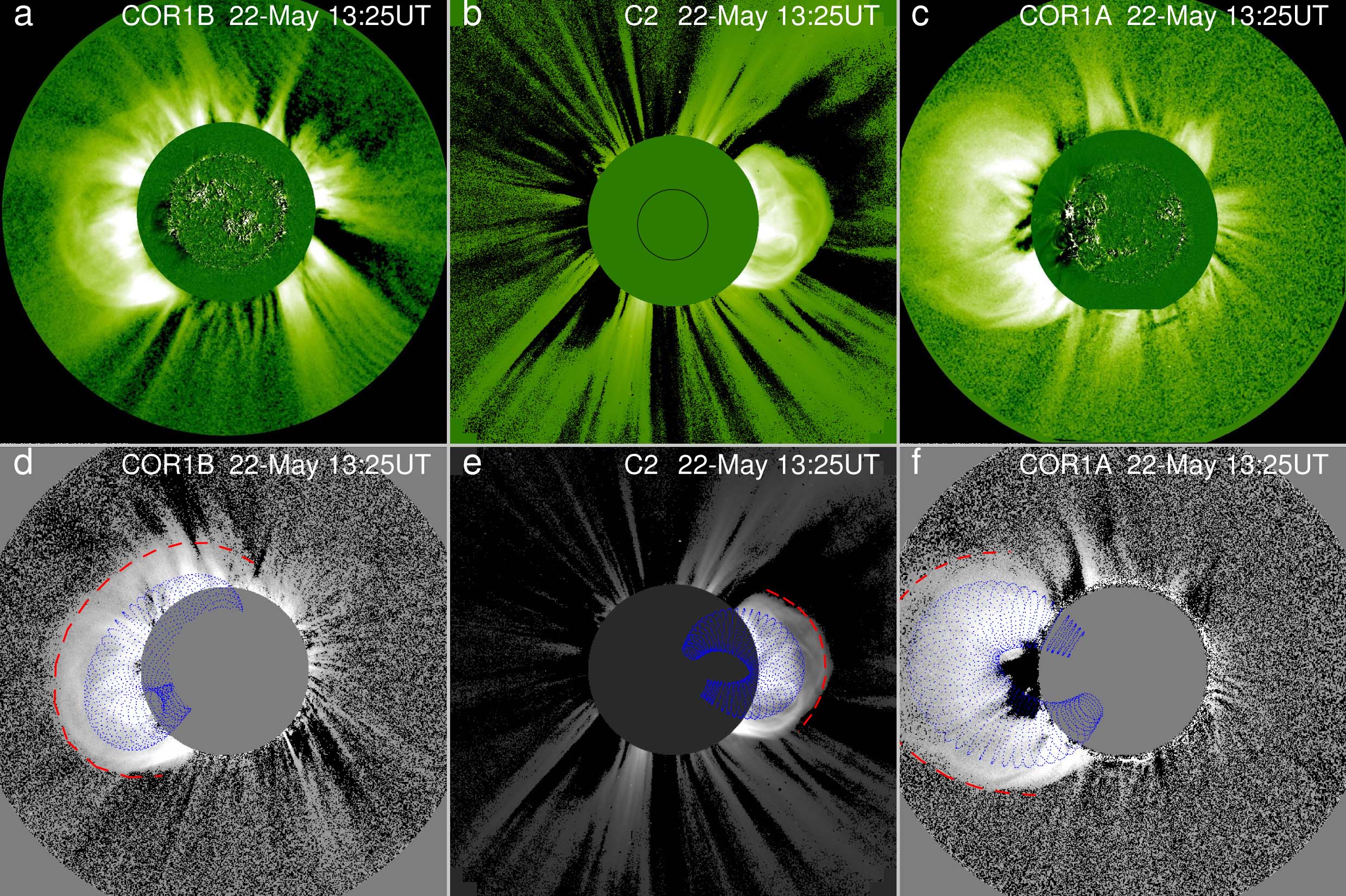}\vspace{-0.0\textwidth}}\\
\caption{(a)--(c) SECCHI/COR1 and LASCO/C2 white-light coronagraph images of the MFR and the compression region ahead. The EUVI 195 {\AA} base difference images are inlaid in (a) and (c). The circle in (b) denotes the location of the solar limb. (d)--(f) GCS modeling of the CME MFR (blue lines). The red dashed lines indicate the shock front.}
(Animations this figure are available in the online journal.)
\label{f4}
\end{figure*}
\subsection{Kinematics of MFR in the Inner Corona}
In order to study the kinematics of the MFR in the inner corona, we take a slice along the eruption direction of the MFR (Figure \ref{f1}b). Figure \ref{f2}a shows the stack plot of the slice in 131 {\AA} passband, in which the heights of the MFR are measured (blue stars). Applying the first order piecewise numerical derivative to the height--time data, we derive the velocity of the MFR (Figure \ref{f2}c). The uncertainty in the velocities arises from the uncertainty in the height measurement, which is estimated to be $\sim$1.7 Mm for AIA observations.

The evolution of the MFR in the inner corona can be described by two kinematic phases: a slow rise phase characterized by a constant velocity and an impulsive acceleration phase characterized by an exponential increase of the velocity \citep[also see;][]{zhang12,cheng13_driver,cheng13_double}. Through fitting a function consisting of linear and exponential components to the height-time measurements of the MFR in the inner corona, \citet{cheng13_double} determined exactly the transition time of the MFR from the first to the second phase, i.e., the onset time of the impulsive acceleration phase. Using the same technique, the acceleration onset of the MFR is derived to be at $\sim$12:31 UT with an uncertainty of 1.5 minutes (green circles in Figure \ref{f2}a and c). The uncertainty is derived by applying 100 Monte Carlo simulations to randomly perturbed time--height data (see Cheng et al. 2013b for details of the technique). 

Figure \ref{f2}a shows that, at $\sim$12:31 UT, the MFR was at a the height of 47$\pm$12 Mm where the decay index $n=$ 1.8$\pm$0.2 of the background magnetic field $B$ is greater than the threshold value 1.5 of the torus instability (Fig. \ref{f2}d), thus most likely triggering the ideal instability \citep{kliem06}, and initiating the impulsive acceleration of the MFR. Here, $B$ is calculated through the potential field model based on the line-of-sight magnetogram on 2013 May 17 (Figure \ref{f2}e) observed by the Helioseismic and Magnetic Imager \citep[HMI;][]{schou12} on board \textit{SDO}. The decay index is calculated by the formula $n$=$-d \ln B/d \ln h$, where $h$ is the height over the solar surface. Note that, the calculation of $B$ and in turn $n$ involves an uncertainty resulting from the fact that the magnetogram was taken 5 days before the eruption. We realize that significant evolution may have occurred between May 17 and 22, however, this is our best option for obtaining this information.

Figure \ref{f2}c displays the velocity evolution of the MFR in the inner corona. During the slow rise phase, the velocity is $\sim$40 km s$^{-1}$. With the impulsive acceleration phase beginning, the velocity increases rapidly from $\sim$40 km s$^{-1}$ to 300 km s$^{-1}$ in 23 minutes, with an average acceleration of $\sim$200 m s$^{-2}$. During this phase, the impulsively accelerating MFR compressed the overlying field and formed a bright front, best visible in the cooler AIA 211 {\AA} passband (Figure \ref{f1}e and Figure \ref{f2}b). One can see that the velocity of the bright front is slightly slower than that of the MFR in the inner corona (Figure \ref{f2}c), indicating that the MFR acts as a driver of the bright front \citep[also see;][]{patsourakos10,cheng12_wave,cheng13_driver}.

\begin{figure*}
\vspace{-0.0\textwidth}
\center {\includegraphics[width=15cm]{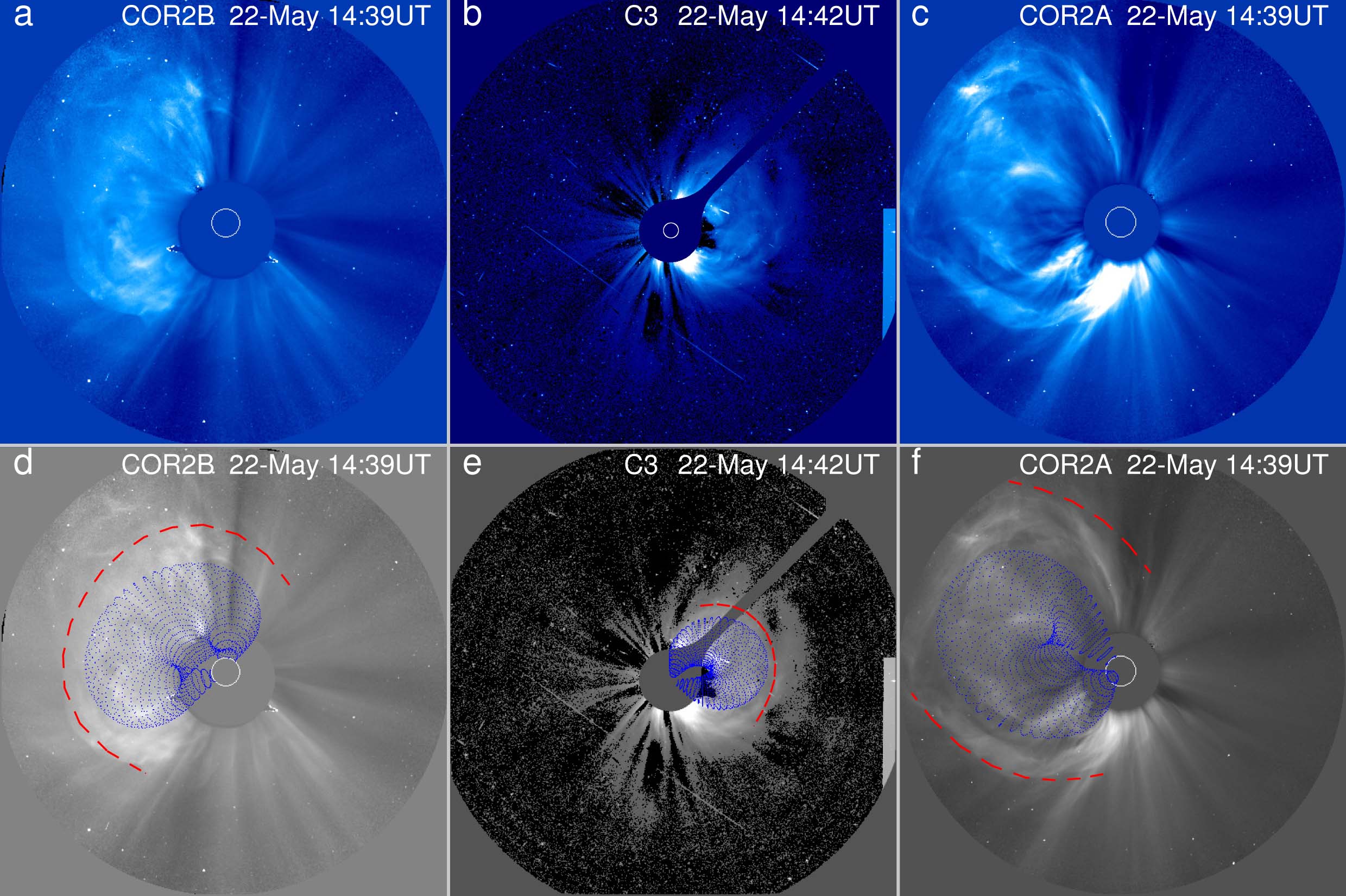}\vspace{-0.0\textwidth}}\\
\caption{(a)--(c) SECCHI/COR2 and LASCO/C3 white-light coronagraph images of the MFR and the compression region ahead. The three circles indicate the location of the solar limb. (d)--(f) GCS modeling of the CME MFR (blue lines). The red dashed lines indicate the shock front.}
(Animations this figure are available in the online journal.)
\label{f5}
\end{figure*}
\subsection{Thermal Property of MFR}

Thanks to the multi-temperature imaging ability of the AIA, we are able to investigate the thermal property of the MFR through differential emission measure (DEM) analysis. Each of the six near-simultaneous AIA EUV images (131, 94, 335, 211, 193, 171 {\AA}) is first processed by the routine ``aia{\_}prep.pro" to the 1.5-level and then the resolution is degraded to 3.6$\arcsec$ by the ``rebin.pro". It guarantees a better coalignment accuracy reducing the error of DEM inversion. Finally, the DEM in each pixel is reconstructed by the routine ``xrt{\_}dem{\_}iterative2.pro" in the SSW package. The code was originally developed by \citet{weber04} and has been modified slightly to work with AIA data \citep{cheng12_dem}. In order to characterize the overall temperature and emission properties of the plasma, we introduce two quantities: the DEM-weighted average temperature $\overline{T}$ and the total emission measure (EM), which are defined as $\overline{T}=\int$DEM($T$)$TdT/ \int$DEM($T$)$dT$ and EM$=\int$DEM($T$)$dT$, respectively.

The evolution of $\overline{T}$ and EM of the eruption region are shown in Figure \ref{f3}. One can see that the quiet background corona is mainly dominated by the plasma with $\overline{T}=$1.5$\sim$2.0 MK and EM$ \approx$10$^{27}$ cm$^{-5}$. In the source AR, an apparent EUV brightening region with $\overline{T}$ $\geq$ 6.0 MK and EM $\geq$ 10$^{28}$ cm$^{-5}$ appears underneath the MFR from 12:15 U towards (Figure \ref{f3}a and \ref{f3}d). As the MFR rises, the brightening expands in the surrounding region. Meanwhile, the two footpoints of the MFR heat to over 8.0 MK, and the corresponding EM increases to over 10$^{28}$ cm$^{-5}$. The results confirm the qualitative argument in Section 2.1 that a slow reconnection may occur and heat the plasma. After $\sim$12:31 UT, with the fast (flare-related) reconnection beginning, the top of the MFR is further heated to $\overline{T}\approx$7.0 MK, while the two footpoints of the MFR are even heated to $\overline{T}\approx$10.0 MK with EM$ \approx$10$^{29}$ cm$^{-5}$. Meanwhile, the flare region is rapidly heated to $\overline{T}\geq$10.0 MK with EM$ \geq$10$^{29}$ cm$^{-5}$. Note that, there is a broad region with $\overline{T}\geq$5.0 MK that is located near the source region of the eruption (outlined by a contour in Figure \ref{f3}a). After inspecting the AIA movies, we find that this high temperature region is due to the previous CME eruption that took place at $\sim$08:00 UT. 

\section{Evolution of MFR in the Outer Corona}

\subsection{White-light Observations of MFR}

In the outer corona, the CME was well observed by the Large Angle and Spectrometric Coronagraph \citep[LASCO;][]{brueckner95} on board the \textit{Solar and Heliospheric Observatory} (\textit{SOHO}) and the Sun-Earth Connection Coronal and Heliospheric Investigation \citep[SECCHI;][]{howard08} on board the \textit{Solar Terrestrial Relations Observatory} (\textit{STEREO-A} and \textit{STEREO-B}). The observations reveal the evolution of the CME from 1.0 to 20.0 R$_\sun$ (Figure \ref{f4} and \ref{f5} and supplementary Movies). 

On 2013 May 22, \textit{STEREO-A} is ahead of the Earth with the separation angle of 138${\degr}$ in the ecliptic plane and \textit{STEREO-B} trails behind the Earth with the separation angle of 143${\degr}$. The three viewpoints provide three distinct projections of the CME. \textbf{The CME changes morphology from a limb to a partial halo progressively from C2 to COR1-A to COR1-B. We can distinguish a coherent bright structure in all three perspectives and a preceding CME front region, best seen in COR2-B. We identify the coherent bright structure as the MFR based on two arguments. First, its orientation is consistent between the AIA and C2 observations. We expect to see it edge-on and hence bright, i.e, no EUV bubble and no dark cavity (Vourlidas et al. 2013 and references therein). Second, the height and velocity of the coherent structure connect smoothly to that of the MFR in the inner corona \citep[also see;][]{maricic04,maricic09}. The CME front region ahead of the MFR appears to consist of three components: plasma pile-up of the MFR, an outer diffuse shock front, and the sheath region between them as expected \citep{vourlidas13}.}

\subsection{GCS modeling of MFR}

Using the graduated cylindrical shell (GCS) model \citep{thernisien06}, we reconstruct the three-dimensional (3D) morphology of the MFR. The model is determined by six parameters: Carrington longitude $\phi$ and latitude $\theta$ of the source region, the tilt angle $\gamma$, the height $r$, and the aspect ratio $\kappa$ of the MFR, as well as the half-angle $\alpha$ between the two legs of the MFR. Taking advantage of the EUVI 195~\AA\ images, we first estimate $\phi$, $\theta$, and $\gamma$ using the location and neutral line of the AR. Then we vary $\alpha$, $\kappa$, and $r$ until we achieve the best visual fit in all three coronagraph images simultaneously. The final positioning and model parameters of the MFR are listed in Table \ref{tb}. The results are displayed in the bottom panels of Figure \ref{f4} and \ref{f5}. It turns out that the GCS model is able to reproduce the 3D morphology of the MFR very well. \textbf{In order to measure the outmost shock, which top corresponds to the top of the CME compression region (depicted by the red dashed lines in Figure \ref{f4} and \ref{f5}), we apply the GCS model to the outer CME envelope and derive the height of the shock \citep[also see;][]{yod12}.}

\begin{figure}
\vspace{-0.0\textwidth}
\center {\includegraphics[width=9cm]{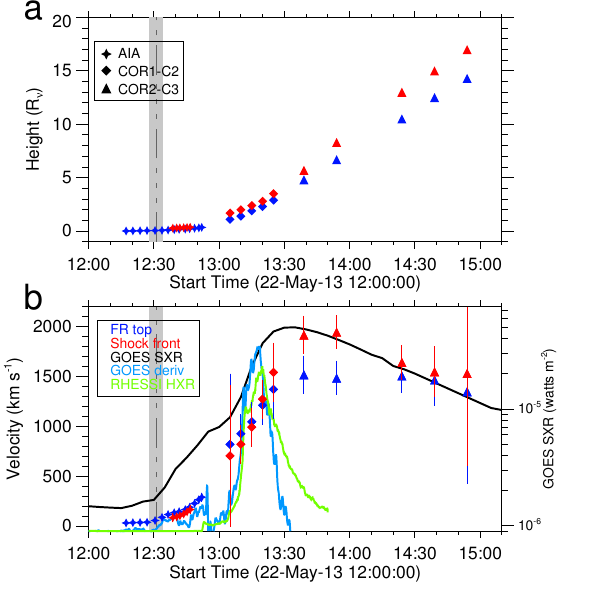}\vspace{-0.0\textwidth}}\\
\caption{Temporal evolution of the heights (a) and velocities (b) of the CME MFR (blue) and shock front (red). The vertical dash-dotted line and the gray region denote the onset time of the impulsive acceleration of the MFR and corresponding uncertainty. The black, blue, and green solid lines show the \textit{GOES} 1--8 {\AA} soft X-ray flux, its time derivative, and the \textit{RHESSI} 15--25 keV hard X-ray flux, respectively. }
\label{f6}
\end{figure}

Figure \ref{f6} shows the height and velocity variations of the MFR and CME shock front. From the height plot, one can see that the MFR and the CME front evolve smoothly from the field of view (FOV) of the AIA into that of the COR1 and COR2. After propagating into the outer corona, both of their heights quickly increase, e.g., from $\sim$1.0 R$_\sun$ at $\sim$13:00 UT to $\sim$15 R$_\sun$ at $\sim$15:00 UT. Whereas, the evolution of the velocity displays different behaviors. From $\sim$13:00 UT to 13:30 UT, the MFR and shock front are impulsively accelerated from $\sim$400 km s$^{-1}$ to $\sim$1500 km s$^{-1}$ with an average acceleration of $\sim$600 m s$^{-2}$. While after $\sim$13:30 UT, the velocities of the MFR and shock front start to decrease. At $\sim$15:00 UT, their velocities decrease to $\sim$1200 km s$^{-1}$. The overall velocity variations of the MFR and shock front are generally consistent with the evolution of the \textit{GOES} soft X-ray (SXR) 1--8 {\AA} flux of the associated flare, i.e., increasing in the rise phase of the flare and decreasing in the decay phase of the flare, as shown in Figure \ref{f6}b. It implies that the two different-scaled eruption phenomena are most probably produced by the same eruption process. 

\section{Summary and Discussions}
\textbf{In this paper, we investigate the MFR eruption on 2013 May 22, including its morphological, kinematical, and thermal properties, and the relationship with the CME in the outer corona. This study might be the most comprehensive study on the MFR to date.} We reveal the evidence for the intrinsic helicity in the MFR: the observed helical threads winding around a possible common axis inside the MFR and the rotational motion of the filament materials along the two legs of the MFR descending into the chromosphere. Further kinematic analysis shows that the whole eruption process of the MFR, including the propagation in the outer corona, can be well characterized by three distinct kinematical evolution phases: the slow rise phase, the impulsive acceleration phase, and the deceleration phase, being similar to that of CMEs \citep{zhang01,zhang06}. 

Based on observations, we interpret the driver behind the slow rise phase as magnetic reconnection in the quasi-separatrix-layers (QSLs) around the MFR. The slow rise phase was initiated when EUV brightenings, underneath the MFR and its two footpoints, started to appear at $\sim$12:00 UT. DEM analysis shows that the MFR top and footpoints were heated to over 6.0 MK. It is conjectured that the heating may result from the reconnection near the periphery of the MFR, i.e., QSLs, where the magnetic linkages undergo drastic changes and hence strong current layers are formed there \citep{demoulin96,savcheva12a,guo13_qmap}. The reconnection in the presumed QSLs has two important roles in the slow rise phase. One is heating the plasma to make the helical threads of the MFR visible in the high temperature passbands of the AIA, e.g., 131 and 94 {\AA}. The other is inferred to be converting the toroidal flux of surrounding sheared fields into the poloidal flux of the MFR. It is because the added poloidal magnetic flux naturally provides an upward magnetic tension to drive the MFR to rise slowly \citep{aulanier10,fan10}.

In order to tentatively uncover the signature of the MFR formation, we inspect the AIA images in six EUV passbands and the line-of-sight magnetograms of the HMI from 2013 May 17 to 22. It is found that the source AR took on a sigmoidal morphology prior to the eruption, as shown in the composite image of AIA 94, 335, and 193 {\AA} (Figure \ref{f2}d). The sigmoid center exhibits the highest temperature, indicating that the reconnection may take place there. The double J-shaped loops that constitute the sigmoid may be from the cooling of reconnected field lines. Moreover, the HMI data indicate that the AR is in its decay phase, in which magnetic cancellation takes place sporadically near the main neutral line. The preliminary observational results seem to imply that the MFR is forming in the corona.

\begin{table*}
\caption{Model and Positioning Parameters of the 2013 May 22 MFR.}
\label{tb}
\begin{tabular}{cccccccc}
\\ \tableline \tableline
  Time     & Lon ($\phi$)   & Lat ($\theta$)  & Tilt ($\gamma$) & H ($r$)       & Ratio ($\alpha$)  & Half-angle ($\kappa$)       \\
  (UT)     & (Deg)          &   (Deg)         & (Deg)           & (R$_\odot$)  &    (r/H)          &   (Deg)          \\
\hline
 & & & Flux Rope &&&\\
 \hline
 13:05        &329              &6.7      &--51        &1.0        &  0.25     &  25         \\
 13:10        &329              &6.7      &--51        &1.4        &  0.25     &  25         \\
 13:15        &329              &6.7      &--51        &1.9        &  0.25     &  30         \\
 13:20        &329              &6.7      &--51        &2.3        &  0.25     &  40         \\
 13:25        &329              &6.7      &--51        &2.9        &  0.25     &  40         \\
 
 13:39        &332              &11.7    &--48        &4.8        &  0.40     &  45         \\
 13:54        &333              &15.0    &--48        &6.7        &  0.40     &  45         \\
 14:24        &333              &15.0    &--48        &10.5      &  0.40     &  45         \\
 14:39        &333              &18.0    &--48        &12.5      &  0.40     &  45         \\
 14:54        &333              &18.0    &--48        &14.3      &  0.40     &  45         \\
 \hline
  & & & Outer CME Envelope &&&\\
 \hline
 13:05        &329              &6.7      &--51        &1.7        &  0.25     &  25         \\
 13:10        &329              &6.7      &--51        &2.0        &  0.25     &  30         \\
 13:15        &329              &6.7      &--51        &2.4        &  0.30     &  30         \\
 13:20        &329              &6.7      &--51        &2.8        &  0.30     &  40         \\
 13:25        &329              &6.7      &--51        &3.5        &  0.35     &  45         \\
 
 13:39        &332              &11.7    &--48        &5.7        &  0.45     &  50         \\
 13:54        &333              &15.0    &--48        &8.3        &  0.45     &  50         \\
 14:24        &333              &18.5    &--48        &13.0      &  0.50     &  60         \\
 14:39        &333              &18.5    &--48        &15.0      &  0.50     &  60         \\
 14:54        &333              &18.5    &--48        &17.0      &  0.50     &  60         \\
\tableline
\vspace{0.03\textwidth}
\end{tabular}
\end{table*}

When the MFR ascends to a height of 47$\pm$12 Mm at $\sim$12:31 UT, it transits into the impulsive acceleration phase. The triggering of the transition is attributed to the torus instability since at that height the decay index, 1.8$\pm$0.2, of the background field is larger the threshold value of torus instability \citep{kliem06,torok05,fan07,aulanier10,savcheva12b,cheng13_double}. As the torus instability commences, the MFR is accelerated upward, stretching the overlying field and forming a current sheet (CS) underneath the MFR. The reconnection in the CS is distinguished from that in the QSLs because of their distinct locations and reconnection rates. Compared to the reconnection in the QSLs, the reconnection in the flare CS has a higher efficiency in accelerating energetic particles, which stream down along the newly-reconnected field lines to produce two well-observed flare ribbons and enhanced flare emission \citep{priest00}. The reconnection in the flare CS also has a higher efficiency in injecting more amount of poloidal flux into the MFR. The added poloidal flux supplies a stronger upward Lorentz self-force to accelerate the MFR. Due to the eruption of the MFR, the magnetic pressure underneath the MFR is reduced, more and more ambient field lines are driven to participate in reconnection so as to further accelerate the FR. Therefore, the reconnection in the flare CS and the MFR eruption are highly coupled in a positive feedback system, which effectively accelerates the MFR and produces the flare emission in the impulsive acceleration phase.

\textbf{Aside from magnetic reconnection, \citet{torok05}, \citet{cheny07}, and \cite{olmedo10} showed that the torus instability itself is capable of driving the MFR eruption. Recently, \citet{patsourakos10} and \citet{patsourakos10_aia} found the lateral overexpansion of the CME bubble after the main flare reconnection phase. \citet{song13} even found out four high-speed CMEs but with very weak flare emission. Both of them are considered to be indirect evidence of the ideal MHD process playing a role in the CME expansion and acceleration.} As for the event we studied, the HXR emission and the time derivative of the SXR emission increased from $\sim$12:31 UT to 13:00 UT slightly if compared to that after $\sim$13:00 UT (Figure \ref{f6}b). We thus suspect that the torus instability is mainly responsible for driving the MFR acceleration in the period of $\sim$12:31 to 13:00 UT. 
  
Another important observation is that we can seamlessly track the MFR from the inner corona to the outer corona. When the MFR propagates into the outer corona, we determine the positions and morphologies of the MFR in the white-light images. Through visually inspecting the whole LASCO CME database, \citet{vourlidas13} recently found that the majority of CMEs in the past solar cycle have a clear MFR structure if excluding the jet and outflow events. The MFR is manifested to be a dark cavity in standard three-component CME events \citep[e.g.,][]{illing85} or a bubble in ``loop"-typed ones \citep[e.g.,][]{patsourakos10}. In our case, the MFR appeared as a coherent bright structure instead the dark cavity. This is probably a result of the projection since we are observing the MFR broadside instead of along its axis \citep{cremades04}. The continuous imaging coverage from the inner to outer corona provides a smooth transition of the EUV channel to the coherent bright structure, which further supports the MFR interpretation. Moreover, identifying the bright structure as the MFR also helps us to understand the properties of the preceding CME front region. The CME bright ``loops" enveloping the MFR probably shows the pile-up of the plasma at the boundary of the MFR and the outer diffuse front denotes the shock front generated by the MFR eruption \citep{vourlidas03,vourlidas13,jin13}.

\acknowledgements We thank the referee, P. F. Chen, P. D{\'e}moulin, Q. Hu, and Y. M. Wang for valuable comments and discussions to improve the manuscript. SDO is a mission of NASAÕs Living With a Star Program, STEREO is the third mission in NASA's Solar Terrestrial Probes program, and SOHO is a mission of international cooperation between ESA and NASA. X.C., Y.G., M.D.D., C.L., and J.Q.S. are supported by NSFC under grants 10933003, 11303016, 11373023, 11203014, and NKBRSF under grants 2011CB811402 and 2014CB744203. J.Z. is supported by NSF grant ATM-0748003, AGS-1156120, and AGS-1249270. A.V. is supported by NASA contract S-136361-Y to the Naval Research Laboratory. O.O. is supported by NASA Living with a Star TR\&T Program NASA NNG11PL10A.


\begin{thebibliography}
\expandafter\ifx\csname natexlab\endcsname\relax\def\natexlab#1{#1}\fi

\bibitem[{{Amari} {et~al.}(2011){Amari}, {Aly}, {Luciani}, {Mikic}, \&
  {Linker}}]{amari11}
{Amari}, T., {Aly}, J.-J., {Luciani}, J.-F., {Mikic}, Z., \& {Linker}, J. 2011,
  \apjl, 742, L27

\bibitem[{{Antiochos} {et~al.}(1999){Antiochos}, {DeVore}, \&
  {Klimchuk}}]{antiochos99}
{Antiochos}, S.~K., {DeVore}, C.~R., \& {Klimchuk}, J.~A. 1999, \apj, 510, 485

\bibitem[{{Aulanier} {et~al.}(2010){Aulanier}, {T{\"o}r{\"o}k}, {D{\'e}moulin},
  \& {DeLuca}}]{aulanier10}
{Aulanier}, G., {T{\"o}r{\"o}k}, T., {D{\'e}moulin}, P., \& {DeLuca}, E.~E.
  2010, \apj, 708, 314

\bibitem[{{Bak-St{\c e}{\'s}licka} {et~al.}(2013){Bak-St{\c e}{\'s}licka},
  {Gibson}, {Fan}, {Bethge}, {Forland}, \& {Rachmeler}}]{bak-steslicka13}
{Bak-St{\c e}{\'s}licka}, U., {Gibson}, S.~E., {Fan}, Y., {Bethge}, C.,
  {Forland}, B., \& {Rachmeler}, L.~A. 2013, \apjl, 770, L28

\bibitem[Berger(2012)]{berger12} Berger, T.\ 2012, Second 
ATST-EAST Meeting: Magnetic Fields from the Photosphere to the Corona., 
463, 147 

\bibitem[{{Brueckner} {et~al.}(1995){Brueckner}, {Howard}, {Koomen},
  {Korendyke}, {et~al.}}]{brueckner95}
{Brueckner}, G.~E., {Howard}, R.~A., {Koomen}, M.~J., {Korendyke}, C.~M.,
  {et~al.} 1995, \solphys, 162, 357

\bibitem[{{Burlaga}(1988)}]{burlaga88}
{Burlaga}, L.~F. 1988, \jgr, 93, 7217

\bibitem[{{Canfield} {et~al.}(1999){Canfield}, {Hudson}, \&
  {McKenzie}}]{canfield99}
{Canfield}, R.~C., {Hudson}, H.~S., \& {McKenzie}, D.~E. 1999, \grl, 26, 627

\bibitem[{{Canou} {et~al.}(2009){Canou}, {Amari}, {Bommier}, {Schmieder},
  {Aulanier}, \& {Li}}]{canou09}
{Canou}, A., {Amari}, T., {Bommier}, V., {Schmieder}, B., {Aulanier}, G., \&
  {Li}, H. 2009, \apjl, 693, L27

\bibitem[{{Chen}(1996)}]{chenj96}
{Chen}, J. 1996, \jgr, 101, 27499

\bibitem[{{Chen}(2011)}]{chen11_review}
{Chen}, P.~F. 2011, Living Reviews in Solar Physics, 8, 1

\bibitem[{{Chen} {et~al.}(2007){Chen}, {Hu}, \& {Sun}}]{cheny07}
{Chen}, Y., {Hu}, Y.~Q., \& {Sun}, S.~J. 2007, \apj, 665, 1421

\bibitem[{{Cheng} {et~al.}(2010){Cheng}, {Ding}, {Guo}, {Zhang}, {Jing}, \&
  {Wiegelmann}}]{cheng10_reflare}
{Cheng}, X., {Ding}, M.~D., {Guo}, Y., {Zhang}, J., {Jing}, J., \&
  {Wiegelmann}, T. 2010, \apjl, 716, L68

\bibitem[{{Cheng} {et~al.}(2013{\natexlab{a}}){Cheng}, {Zhang}, {Ding}, {Liu},
  \& {Poomvises}}]{cheng13_driver}
{Cheng}, X., {Zhang}, J., {Ding}, M.~D., {Liu}, Y., \& {Poomvises}, W.
  2013{\natexlab{a}}, \apj, 763, 43

\bibitem[{{Cheng} {et~al.}(2013{\natexlab{b}}){Cheng}, {Zhang}, {Ding},
  {Olmedo}, {Sun}, {Guo}, \& {Liu}}]{cheng13_double}
{Cheng}, X., {Zhang}, J., {Ding}, M.~D., {Olmedo}, O., {Sun}, X.~D., {Guo}, Y.,
  \& {Liu}, Y. 2013{\natexlab{b}}, \apjl, 769, L25

\bibitem[{{Cheng} {et~al.}(2011){Cheng}, {Zhang}, {Liu}, \&
  {Ding}}]{cheng11_fluxrope}
{Cheng}, X., {Zhang}, J., {Liu}, Y., \& {Ding}, M.~D. 2011, \apjl, 732, L25

\bibitem[{{Cheng} {et~al.}(2012{\natexlab{a}}){Cheng}, {Zhang}, {Olmedo},
  {Vourlidas}, {Ding}, \& {Liu}}]{cheng12_wave}
{Cheng}, X., {Zhang}, J., {Olmedo}, O., {Vourlidas}, A., {Ding}, M.~D., \&
  {Liu}, Y. 2012{\natexlab{a}}, \apjl, 745, L5

\bibitem[{{Cheng} {et~al.}(2012{\natexlab{b}}){Cheng}, {Zhang}, {Saar}, \&
  {Ding}}]{cheng12_dem}
{Cheng}, X., {Zhang}, J., {Saar}, S.~H., \& {Ding}, M.~D. 2012{\natexlab{b}},
  \apj, 761, 62

\bibitem[Cremades 
\& Bothmer(2004)]{cremades04} Cremades, H., \& Bothmer, V.\ 2004, \aap, 422, 307 

\bibitem[{{D{\'e}moulin} \& {Aulanier}(2010)}]{demoulin10}
{D{\'e}moulin}, P., \& {Aulanier}, G. 2010, \apj, 718, 1388

\bibitem[{{Demoulin} {et~al.}(1996){Demoulin}, {Henoux}, {Priest}, \&
  {Mandrini}}]{demoulin96}
{Demoulin}, P., {Henoux}, J.~C., {Priest}, E.~R., \& {Mandrini}, C.~H. 1996,
  \aap, 308, 643

\bibitem[{{Dove} {et~al.}(2011){Dove}, {Gibson}, {Rachmeler}, {Tomczyk}, \&
  {Judge}}]{dove11}
{Dove}, J.~B., {Gibson}, S.~E., {Rachmeler}, L.~A., {Tomczyk}, S., \& {Judge},
  P. 2011, \apjl, 731, L1

\bibitem[{{Fan}(2001)}]{fan01}
{Fan}, Y. 2001, \apjl, 554, L111

\bibitem[{{Fan}(2010)}]{fan10}
---. 2010, \apj, 719, 728

\bibitem[{{Fan} \& {Gibson}(2007)}]{fan07}
{Fan}, Y., \& {Gibson}, S.~E. 2007, \apj, 668, 1232

\bibitem[{{Forbes} \& {Isenberg}(1991)}]{forbes91}
{Forbes}, T.~G., \& {Isenberg}, P.~A. 1991, \apj, 373, 294

\bibitem[Gibson et al.(2004)]{gibson04} Gibson, S.~E., Fan, Y., 
Mandrini, C., Fisher, G., \& Demoulin, P.\ 2004, \apj, 617, 600 


\bibitem[{{Gibson} \& {Fan}(2006)}]{gibson06_jgr}
{Gibson}, S.~E., \& {Fan}, Y. 2006, Journal of Geophysical Research (Space
  Physics), 111, 12103

\bibitem[{{Gibson} {et~al.}(2006){Gibson}, {Foster}, {Burkepile}, {de Toma}, \&
  {Stanger}}]{gibson06_apj}
{Gibson}, S.~E., {Foster}, D., {Burkepile}, J., {de Toma}, G., \& {Stanger}, A.
  2006, \apj, 641, 590

\bibitem[{{Gosling}(1993)}]{gosling93}
{Gosling}, J.~T. 1993, \jgr, 98, 18937

\bibitem[{{Green} \& {Kliem}(2009)}]{green09}
{Green}, L.~M., \& {Kliem}, B. 2009, \apjl, 700, L83

\bibitem[{{Guo} \& {Wu}(1998)}]{guo98}
{Guo}, W.~P., \& {Wu}, S.~T. 1998, \apj, 494, 419

\bibitem[{{Guo} {et~al.}(2013){Guo}, {Ding}, {Cheng}, {Zhao}, \&
  {Pariat}}]{guo13_qmap}
{Guo}, Y., {Ding}, M.~D., {Cheng}, X., {Zhao}, J., \& {Pariat}, E. 2013,
  \apjsub

\bibitem[{{Guo} {et~al.}(2010){Guo}, {Schmieder}, {D{\'e}moulin}, {Wiegelmann},
  {Aulanier}, {T{\"o}r{\"o}k}, \& {Bommier}}]{guo10_filament}
{Guo}, Y., {Schmieder}, B., {D{\'e}moulin}, P., {Wiegelmann}, T., {Aulanier},
  G., {T{\"o}r{\"o}k}, T., \& {Bommier}, V. 2010, \apj, 714, 343

\bibitem[{{Howard} {et~al.}(2008){Howard}, {Moses}, {Vourlidas}, {Newmark},
  {Socker}, {Plunkett}, {Korendyke}, {Cook}, {Hurley}, {Davila}, {Thompson},
  {St Cyr}, {Mentzell}, {Mehalick}, {Lemen}, {Wuelser}, {Duncan}, {Tarbell},
  {Wolfson}, {Moore}, {Harrison}, {Waltham}, {Lang}, {Davis}, {Eyles},
  {Mapson-Menard}, {Simnett}, {Halain}, {Defise}, {Mazy}, {Rochus}, {Mercier},
  {Ravet}, {Delmotte}, {Auchere}, {Delaboudiniere}, {Bothmer}, {Deutsch},
  {Wang}, {Rich}, {Cooper}, {Stephens}, {Maahs}, {Baugh}, {McMullin}, \&
  {Carter}}]{howard08}
{Howard}, R.~A., {et~al.} 2008, \ssr, 136, 67

\bibitem[{{Hughes} \& {Sibeck}(1987)}]{hughes87}
{Hughes}, W.~J., \& {Sibeck}, D.~G. 1987, \grl, 14, 636

\bibitem[{{Illing} \& {Hundhausen}(1985)}]{illing85}
{Illing}, R.~M.~E., \& {Hundhausen}, A.~J. 1985, \jgr, 90, 275

\bibitem[{{Inoue} {et~al.}(2013){Inoue}, {Hayashi}, {Shiota}, {Magara}, \&
  {Choe}}]{inoue13}
{Inoue}, S., {Hayashi}, K., {Shiota}, D., {Magara}, T., \& {Choe}, G.~S. 2013,
  \apj, 770, 79

\bibitem[{{Jiang} {et~al.}(2013){Jiang}, {Feng}, {Wu}, \& {Hu}}]{jiang13}
{Jiang}, C., {Feng}, X., {Wu}, S.~T., \& {Hu}, Q. 2013, \apjl, 771, L30

\bibitem[{{Jin} {et~al.}(2013){Jin}, {Manchester}, {van der Holst}, {Oran},
  {Sokolov}, {Toth}, {Liu}, {Sun}, \& {Gombosi}}]{jin13}
{Jin}, M., {et~al.} 2013, \apj, 773, 50

\bibitem[{{Karpen} {et~al.}(2012){Karpen}, {Antiochos}, \& {DeVore}}]{karpen12}
{Karpen}, J.~T., {Antiochos}, S.~K., \& {DeVore}, C.~R. 2012, \apj, 760, 81

\bibitem[{{Kliem} \& {T{\"o}r{\"o}k}(2006)}]{kliem06}
{Kliem}, B., \& {T{\"o}r{\"o}k}, T. 2006, \prl, 96, 255002

\bibitem[{{Lemen} {et~al.}(2012){Lemen}, {Title}, {Akin}, {Boerner}, {Chou},
  {Drake}, {Duncan}, {Edwards}, {Friedlaender}, {Heyman}, {Hurlburt}, {Katz},
  {Kushner}, {Levay}, {Lindgren}, {Mathur}, {McFeaters}, {Mitchell}, {Rehse},
  {Schrijver}, {Springer}, {Stern}, {Tarbell}, {Wuelser}, {Wolfson}, {Yanari},
  {Bookbinder}, {Cheimets}, {Caldwell}, {Deluca}, {Gates}, {Golub}, {Park},
  {Podgorski}, {Bush}, {Scherrer}, {Gummin}, {Smith}, {Auker}, {Jerram},
  {Pool}, {Soufli}, {Windt}, {Beardsley}, {Clapp}, {Lang}, \&
  {Waltham}}]{lemen12}
{Lemen}, J.~R., {et~al.} 2012, \solphys, 275, 17

\bibitem[{{Lepping} {et~al.}(1990){Lepping}, {Burlaga}, \& {Jones}}]{lepping90}
{Lepping}, R.~P., {Burlaga}, L.~F., \& {Jones}, J.~A. 1990, \jgr, 95, 11957

\bibitem[{{Li} \& {Zhang}(2013{\natexlab{a}})}]{lileping13}
{Li}, L.~P., \& {Zhang}, J. 2013{\natexlab{a}}, \aap, 552, L11

\bibitem[{{Li} \& {Zhang}(2013{\natexlab{b}})}]{liting13}
{Li}, T., \& {Zhang}, J. 2013{\natexlab{b}}, \apjl, 770, L25

\bibitem[{{Li} {et~al.}(2012){Li}, {Morgan}, {Leonard}, \& {Jeska}}]{lixing12}
{Li}, X., {Morgan}, H., {Leonard}, D., \& {Jeska}, L. 2012, \apjl, 752, L22

\bibitem[{{Liu} {et~al.}(2010{\natexlab{a}}){Liu}, {Liu}, {Wang}, {Deng}, \&
  {Wang}}]{liur10}
{Liu}, R., {Liu}, C., {Wang}, S., {Deng}, N., \& {Wang}, H. 2010{\natexlab{a}},
  \apjl, 725, L84

\bibitem[{{Liu} {et~al.}(2010{\natexlab{b}}){Liu}, {Davies}, {Luhmann},
  {Vourlidas}, {Bale}, \& {Lin}}]{liuy10}
{Liu}, Y., {Davies}, J.~A., {Luhmann}, J.~G., {Vourlidas}, A., {Bale}, S.~D.,
  \& {Lin}, R.~P. 2010{\natexlab{b}}, \apjl, 710, L82

\bibitem[{{Liu} {et~al.}(2011){Liu}, {Luhmann}, {Bale}, \& {Lin}}]{liuying11}
{Liu}, Y., {Luhmann}, J.~G., {Bale}, S.~D., \& {Lin}, R.~P. 2011, \apj, 734, 84

\bibitem[{{Liu} {et~al.}(2008){Liu}, {Luhmann}, {Huttunen}, {Lin}, {Bale},
  {Russell}, \& {Galvin}}]{liuy08}
{Liu}, Y., {Luhmann}, J.~G., {Huttunen}, K.~E.~J., {Lin}, R.~P., {Bale}, S.~D.,
  {Russell}, C.~T., \& {Galvin}, A.~B. 2008, \apjl, 677, L133

\bibitem[{{Low} \& {Hundhausen}(1995)}]{low95_apj}
{Low}, B.~C., \& {Hundhausen}, J.~R. 1995, \apj, 443, 818

\bibitem[{{Lynch} {et~al.}(2008){Lynch}, {Antiochos}, {DeVore}, {Luhmann}, \&
  {Zurbuchen}}]{lynch08}
{Lynch}, B.~J., {Antiochos}, S.~K., {DeVore}, C.~R., {Luhmann}, J.~G., \&
  {Zurbuchen}, T.~H. 2008, \apj, 683, 1192

\bibitem[{{Mackay} {et~al.}(2010){Mackay}, {Karpen}, {Ballester}, {Schmieder},
  \& {Aulanier}}]{mackay10}
{Mackay}, D.~H., {Karpen}, J.~T., {Ballester}, J.~L., {Schmieder}, B., \&
  {Aulanier}, G. 2010, \ssr, 151, 333

\bibitem[{{Magara}(2006)}]{magara06}
{Magara}, T. 2006, \apj, 653, 1499

\bibitem[{{Manchester} {et~al.}(2004){Manchester}, {Gombosi}, {DeZeeuw}, \&
  {Fan}}]{manchester04}
{Manchester}, IV, W., {Gombosi}, T., {DeZeeuw}, D., \& {Fan}, Y. 2004, \apj,
  610, 588

\bibitem[{{Mari{\v c}i{\'c}} {et~al.}(2009){Mari{\v c}i{\'c}}, {Vr{\v s}nak},
  \& {Ro{\v s}a}}]{maricic09}
{Mari{\v c}i{\'c}}, D., {Vr{\v s}nak}, B., \& {Ro{\v s}a}, D. 2009, \solphys,
  260, 177

\bibitem[{{Mari{\v c}i{\'c}} {et~al.}(2004){Mari{\v c}i{\'c}}, {Vr{\v s}nak},
  {Stanger}, \& {Veronig}}]{maricic04}
{Mari{\v c}i{\'c}}, D., {Vr{\v s}nak}, B., {Stanger}, A.~L., \& {Veronig}, A.
  2004, \solphys, 225, 337

\bibitem[{{McKenzie} \& {Canfield}(2008)}]{mcKenzie08}
{McKenzie}, D.~E., \& {Canfield}, R.~C. 2008, \aap, 481, L65

\bibitem[{{Meier} {et~al.}(2001){Meier}, {Koide}, \& {Uchida}}]{meier01}
{Meier}, D.~L., {Koide}, S., \& {Uchida}, Y. 2001, Science, 291, 84

\bibitem[{{Moldwin} \& {Hughes}(1991)}]{moldwin91}
{Moldwin}, M.~B., \& {Hughes}, W.~J. 1991, \jgr, 96, 14051

\bibitem[{{Moore} {et~al.}(2001){Moore}, {Sterling}, {Hudson}, \&
  {Lemen}}]{moore01}
{Moore}, R.~L., {Sterling}, A.~C., {Hudson}, H.~S., \& {Lemen}, J.~R. 2001,
  \apj, 552, 833

\bibitem[{{Morris} {et~al.}(2006){Morris}, {Uchida}, \& {Do}}]{morris06}
{Morris}, M., {Uchida}, K., \& {Do}, T. 2006, \nat, 440, 308

\bibitem[{{O'Dwyer} {et~al.}(2010){O'Dwyer}, {Del Zanna}, {Mason}, {Weber}, \&
  {Tripathi}}]{odwyer10}
{O'Dwyer}, B., {Del Zanna}, G., {Mason}, H.~E., {Weber}, M.~A., \& {Tripathi},
  D. 2010, \aap, 521, A21

\bibitem[{{Okamoto} {et~al.}(2008){Okamoto}, {Tsuneta}, {Lites}, {Kubo},
  {Yokoyama}, {Berger}, {Ichimoto}, {Katsukawa}, {Nagata}, {Shibata},
  {Shimizu}, {Shine}, {Suematsu}, {Tarbell}, \& {Title}}]{okamoto08}
{Okamoto}, T.~J., {et~al.} 2008, \apjl, 673, L215

\bibitem[{{Olmedo} \& {Zhang}(2010)}]{olmedo10}
{Olmedo}, O., \& {Zhang}, J. 2010, \apj, 718, 433

\bibitem[{{Patsourakos} \& {Vourlidas}(2012)}]{patsourakos12}
{Patsourakos}, S., \& {Vourlidas}, A. 2012, \solphys, 281, 187

\bibitem[{{Patsourakos} {et~al.}(2010){Patsourakos}, {Vourlidas}, \&
  {Kliem}}]{patsourakos10}
{Patsourakos}, S., {Vourlidas}, A., \& {Kliem}, B. 2010, \aap, 522, A100

\bibitem[Patsourakos et al.(2010)]{patsourakos10_aia} Patsourakos, S., 
Vourlidas, A., \& Stenborg, G.\ 2010, \apjl, 724, L188 

\bibitem[{{Patsourakos} {et~al.}(2013){Patsourakos}, {Vourlidas}, \&
  {Stenborg}}]{patsourakos13}
{Patsourakos}, S., {Vourlidas}, A., \& {Stenborg}, G. 2013, \apj, 764, 125

\bibitem[Poomvises et al.(2012)]{yod12} Poomvises, W., 
Gopalswamy, N., Yashiro, S., Kwon, R.-Y., 
\& Olmedo, O.\ 2012, \apj, 758, 118 

\bibitem[{{Priest} \& {Forbes}(2000)}]{priest00}
{Priest}, E., \& {Forbes}, T. 2000, {Magnetic Reconnection}, ed. {Priest, E.~\&
  Forbes, T.} (Cambridge University Press)

\bibitem[{{R{\'e}gnier} {et~al.}(2011){R{\'e}gnier}, {Walsh}, \&
  {Alexander}}]{regnier11}
{R{\'e}gnier}, S., {Walsh}, R.~W., \& {Alexander}, C.~E. 2011, \aap, 533, L1

\bibitem[{{Russell} \& {Elphic}(1979)}]{russell79}
{Russell}, C.~T., \& {Elphic}, R.~C. 1979, \nat, 279, 616

\bibitem[{{Savcheva} {et~al.}(2012{\natexlab{a}}){Savcheva}, {Pariat}, {van
  Ballegooijen}, {Aulanier}, \& {DeLuca}}]{savcheva12b}
{Savcheva}, A., {Pariat}, E., {van Ballegooijen}, A., {Aulanier}, G., \&
  {DeLuca}, E. 2012{\natexlab{a}}, \apj, 750, 15

\bibitem[{{Savcheva} {et~al.}(2012{\natexlab{b}}){Savcheva}, {van
  Ballegooijen}, \& {DeLuca}}]{savcheva12a}
{Savcheva}, A.~S., {van Ballegooijen}, A.~A., \& {DeLuca}, E.~E.
  2012{\natexlab{b}}, \apj, 744, 78

\bibitem[{{Schou} {et~al.}(2012){Schou}, {Scherrer}, {Bush}, {Wachter},
  {Couvidat}, {Rabello-Soares}, {Bogart}, {Hoeksema}, {Liu}, {Duvall}, {Akin},
  {Allard}, {Miles}, {Rairden}, {Shine}, {Tarbell}, {Title}, {Wolfson},
  {Elmore}, {Norton}, \& {Tomczyk}}]{schou12}
{Schou}, J., {et~al.} 2012, \solphys, 275, 229

\bibitem[{{Song} {et~al.}(2013){Song}, {Chen}, {Ye}, {Han}, {Du}, {Li},
  {Zhang}, \& {Hu}}]{song13}
{Song}, H.~Q., {Chen}, Y., {Ye}, D.~D., {Han}, G.~Q., {Du}, G.~H., {Li}, G.,
  {Zhang}, J., \& {Hu}, Q. 2013, \apj, 773, 129

\bibitem[{{Su} {et~al.}(2011){Su}, {Surges}, {van Ballegooijen}, {DeLuca}, \&
  {Golub}}]{suyingna11}
{Su}, Y., {Surges}, V., {van Ballegooijen}, A., {DeLuca}, E., \& {Golub}, L.
  2011, \apj, 734, 53

\bibitem[{{Su} \& {van Ballegooijen}(2012)}]{suyingna12}
{Su}, Y., \& {van Ballegooijen}, A. 2012, \apj, 757, 168

\bibitem[{{Thernisien} {et~al.}(2006){Thernisien}, {Howard}, \&
  {Vourlidas}}]{thernisien06}
{Thernisien}, A.~F.~R., {Howard}, R.~A., \& {Vourlidas}, A. 2006, \apj, 652,
  763

\bibitem[{{Titov} \& {D{\'e}moulin}(1999)}]{titov99}
{Titov}, V.~S., \& {D{\'e}moulin}, P. 1999, \aap, 351, 707

\bibitem[{{T{\"o}r{\"o}k} \& {Kliem}(2005)}]{torok05}
{T{\"o}r{\"o}k}, T., \& {Kliem}, B. 2005, \apjl, 630, L97

\bibitem[{{T{\"o}r{\"o}k} {et~al.}(2004){T{\"o}r{\"o}k}, {Kliem}, \&
  {Titov}}]{torok04}
{T{\"o}r{\"o}k}, T., {Kliem}, B., \& {Titov}, V.~S. 2004, \aap, 413, L27

\bibitem[{{Tripathi} {et~al.}(2009){Tripathi}, {Kliem}, {Mason}, {Young}, \&
  {Green}}]{tripathi09}
{Tripathi}, D., {Kliem}, B., {Mason}, H.~E., {Young}, P.~R., \& {Green}, L.~M.
  2009, \apjl, 698, L27

\bibitem[{{Vourlidas} {et~al.}(2013){Vourlidas}, {Lynch}, {Howard}, \&
  {Li}}]{vourlidas13}
{Vourlidas}, A., {Lynch}, B.~J., {Howard}, R.~A., \& {Li}, Y. 2013, \solphys,
  284, 179

\bibitem[{{Vourlidas} {et~al.}(2003){Vourlidas}, {Wu}, {Wang}, {Subramanian},
  \& {Howard}}]{vourlidas03}
{Vourlidas}, A., {Wu}, S.~T., {Wang}, A.~H., {Subramanian}, P., \& {Howard},
  R.~A. 2003, \apj, 598, 1392

\bibitem[Wang 
\& Stenborg(2010)]{wangym10} Wang, Y.-M., \& Stenborg, G.\ 2010, \apjl, 719, L181 

\bibitem[{{Weber} {et~al.}(2004){Weber}, {Deluca}, {Golub}, \&
  {Sette}}]{weber04}
{Weber}, M.~A., {Deluca}, E.~E., {Golub}, L., \& {Sette}, A.~L. 2004, in IAU
  Symposium, Vol. 223, Multi-Wavelength Investigations of Solar Activity, ed.
  A.~V. {Stepanov}, E.~E. {Benevolenskaya}, \& A.~G. {Kosovichev}, 321--328

\bibitem[{{Wood} {et~al.}(1999){Wood}, {Karovska}, {Chen}, {Brueckner}, {Cook},
  \& {Howard}}]{wood99}
{Wood}, B.~E., {Karovska}, M., {Chen}, J., {Brueckner}, G.~E., {Cook}, J.~W.,
  \& {Howard}, R.~A. 1999, \apj, 512, 484

\bibitem[{{Yan} {et~al.}(2001){Yan}, {Deng}, {Karlick{\'y}}, {Fu}, {Wang}, \&
  {Liu}}]{yan01}
{Yan}, Y., {Deng}, Y., {Karlick{\'y}}, M., {Fu}, Q., {Wang}, S., \& {Liu}, Y.
  2001, \apjl, 551, L115

\bibitem[{{Yuan} {et~al.}(2009){Yuan}, {Lin}, {Wu}, \& {Ho}}]{yuan09}
{Yuan}, F., {Lin}, J., {Wu}, K., \& {Ho}, L.~C. 2009, \mnras, 395, 2183

\bibitem[{{Zhang} {et~al.}(2012){Zhang}, {Cheng}, \& {Ding}}]{zhang12}
{Zhang}, J., {Cheng}, X., \& {Ding}, M.-D. 2012, Nature Communications, 3, 747

\bibitem[{{Zhang} \& {Dere}(2006)}]{zhang06}
{Zhang}, J., \& {Dere}, K.~P. 2006, \apj, 649, 1100

\bibitem[{{Zhang} {et~al.}(2001){Zhang}, {Dere}, {Howard}, {Kundu}, \&
  {White}}]{zhang01}
{Zhang}, J., {Dere}, K.~P., {Howard}, R.~A., {Kundu}, M.~R., \& {White}, S.~M.
  2001, \apj, 559, 452

\bibitem[{{Zhang} {et~al.}(2007){Zhang}, {Richardson}, {Webb}, {Gopalswamy},
  {Huttunen}, {Kasper}, {Nitta}, {Poomvises}, {Thompson}, {Wu}, {Yashiro}, \&
  {Zhukov}}]{zhang07_icme}
{Zhang}, J., {et~al.} 2007, Journal of Geophysical Research (Space Physics),
  112, 10102

\end{thebibliography}

\end{document}